# Determining the utility of ultrafast nonlinear contrast enhanced and super resolution ultrasound for imaging microcirculation in the human small intestine.

Clotilde Vié[1], Martina Tashkova[2,3], James Burn[2], Matthieu Toulemonde[1], Jipeng Yan[1], Jingwen Zhu[1], Cameron A. B. Smith[1], Biao Huang[1], Su Yan[1], Kevin G. Murphy*[3], Gary Frost*[3], Meng-Xing Tang*[1].

[1] Department of Biomedical Engineering, Imperial College London, UK
[2] Imperial College Healthcare NHS Trust, UK
[3] Department of Metabolism, Digestion and Reproduction, Imperial College London, UK

*Seniors authors with equal participation.

## ABSTRACT

**Background** The regulation of intestinal blood flow is critical to gastrointestinal function. Imaging the intestinal mucosal micro-circulation *in vivo* has the potential to provide new insight into the gut physiology and pathophysiology.

**Objective** We aimed to determine whether ultrafast contrast enhanced ultrasound (CEUS) and super-resolution ultrasound localisation microscopy (SRUS/ULM) could be a useful tool for imaging the small intestine microcirculation *in vivo* non-invasively and for detecting changes in blood flow in the duodenum.

**Design** Ultrafast CEUS and SRUS/ULM were used to image the small intestinal microcirculation in a cohort of 20 healthy volunteers (BMI<25). Participants were imaged while conscious and either having been fasted, or following ingestion of a liquid meal or water control, or under acute stress.

**Results** For the first time we have performed ultrafast CEUS and ULM on the human small intestine, providing unprecedented resolution images of the intestinal microcirculation. We evaluated flow speed inside small vessels in healthy volunteers (2·78 ± 0·05 mm/s, mean ± SEM) and quantified changes in the perfusion of this microcirculation in response to nutrient ingestion. Perfusion of the microvasculature of the intestinal mucosa significantly increased post-prandially (36·2% ± 12·2%, mean ± SEM, $p<0·05$). The feasibility of 3D SRUS/ULM was also demonstrated.

**Conclusion** This study demonstrates the potential utility of ultrafast CEUS for assessing perfusion and detecting changes in blood flow in the duodenum. SRUS/ULM also proved a useful tool to image the microvascular blood flow *in vivo* non-invasively and to evaluate blood speed inside the microvasculature of the human small intestine.

**INTRODUCTION**

The small intestine is the major site of food digestion and nutrient absorption and plays an important role in appetite and systemic metabolism though the release of gut hormones and neuronal signalling via the vagus nerve[1,2]. Gastrointestinal blood flow is critically important to the function and integrity of the gastrointestinal tract and imaging intestinal blood flow in humans has the potential to provide insight into gut physiology and pathophysiology, including inflammatory bowel disease and metabolic disease. However, because accessing and visualising gastrointestinal microvascular flow is difficult, we currently know little about how it is regulated and how it changes in disease.

Characterising dynamic changes in microvascular blood flow *in vivo* is challenging, particularly in tissues deeper than 1cm. Today, the evaluation of the intestinal microcirculation is mainly done invasively by confocal laser endomicroscopy (CLE). Commonly used non-invasive clinical imaging approaches such as magnetic resonance imaging (MRI) and computed tomography (CT) lack the necessary spatial resolution or sensitivity, while optical imaging is confined to superficial vessels. Contrast enhanced ultrasound (CEUS) uses the intravenous administration of microbubble contrast agents, and is well established to for characterising indeterminate liver lesions, as a second or third modality after triple phase contrast MRI and/or CT, though it has not previously been used to image the human intestine. Super-resolution ultrasound localization microscopy (SRUS/ULM), based on CEUS acquisitions, was demonstrated *in vivo* in mice and rats, and has proven a powerful tool to observe microvascular flow, including in deeper tissues[3,4]. After imaging and processing, individual microbubbles (MBs) are localised and tracked, allowing for the reconstruction of the vascular network[5,6]. Imaging of the bubbles results in a point-spread function that is reduced to a single point when localising its centre, making it possible to overcome the diffraction limit and non-invasively image the microcirculation in conscious humans. Mapping the microcirculation of vessels with diameters of a few hundred microns or less has been demonstrated in human tissues including the lower limb[7], the brain[8], the liver[9], the lymph node[10], the glomerulus of the kidney[11], the myocardium of the heart[12], and in breast cancer[13,14]. However, imaging of the human gut microcirculation, which has potentially important applications in the investigation of gut physiology and the diagnosis and monitoring of gastrointestinal disease, has not previously been explored.

We investigated the potential utility of ultrafast CEUS and SRUS/ULM in imaging the human intestinal microcirculation. Here we show the first super-resolution (SR) images of the human intestine and demonstrate that we can quantitate the intestinal microvascular response to nutrient ingestion using CEUS. These data suggest that ultrafast CEUS and SRUS/ULM may have utility for investigating gastrointestinal physiology and disease.

## METHODS

### Ethical approval

The study was granted research ethical approval (Fulham NHS Health Research Ethics Committee, UK; 19/LO/1766 IRAS: 254475) and conducted in accordance with the Declaration of Helsinki. Health Research Authority (HRA) and Health and Care Research Wales (HCRW) approval was obtained prior to starting the study. All participants provided written informed consent prior to study commencement. Contrast enhanced imaging was performed on healthy volunteers at the Imperial Clinical Research Facility (ICRF) at Hammersmith Hospital between February 2024 and September 2024.

### Recruitment

A cohort of 20 healthy participants was recruited. Inclusion criteria: male or female, aged between 18 and 50, with a Body Mass Index (BMI) < 25 and no evidence of insulin resistance/diabetes, and willing to consent to medical notes and diagnostic test results being reviewed, captured, and recorded by the clinical research study team. Exclusion criteria: fistulizing Crohn's disease; severe cardiac, respiratory or renal disorders; abdominal surgery in the last 8 weeks; pregnancy or lactation; smoking; previous allergic reaction to SonoVue components; use of any medications except for paracetamol and oral contraception.

### Studies and Protocols

We conducted a series of first in human investigative studies. Specific studies were conducted to investigate the ability of ULM to detect the impact of the nutritional content and volume of a liquid meal on blood flow in the small intestine. An additional pilot study was carried out to investigate the effects of mild stress. Twenty healthy volunteers were asked to fast for at least 6 hours prior to each imaging session to reduce the gut motion induced by digestion. For each of the visits at the clinical investigation unit, volunteers were injected with 3 separate boluses of 1·6 ml of Sonovue microbubbles and multiple contrast enhanced ultrasound (CEUS) acquisitions taken. Each acquisition lasted 5 seconds, and volunteers were asked to hold their breath during this time, to limit breathing motion.

To facilitate the imaging of comparable locations between volunteers, section D3 of the duodenum was chosen for the imaging. This 10-15 cm long horizontal section travels laterally to the left, crossing the vertebral column, the abdominal aorta and the inferior vena cava, terminating at the inferior duodenal flexure (Figure 1(b)-(f)). In addition, sections of the jejunum, which has the advantage of often lying just below the abdominal muscles, were also imaged, typically providing higher resolution images, but also images more likely to be limited by gut motion. Those jejunal images with little motion were further processed to generate SR images and visualise the microcirculation.

**Using ultrasound to investigate changes in blood supply to the small intestine following food ingestion.** The oxygen and energy demand of the small intestine increase following the commencement of a meal, driving increased blood flow to provide fuel and to transport the absorbed products of digestion to other tissues[1,2,15]. To determine the utility of ultrasound in measuring the response of the small intestinal microvasculature to food, participants attended the clinical investigation unit twice. On their first visit, 12 participants were imaged while being fasted and then twice either 5 min, 10 min, 20 min or 30 min after the ingestion of a 200 ml Ensure® Plus Vanilla Milkshake (Abbott Laboratories, USA, 300 kcal, 12.5g of protein (49·98 kcal), 40·4g of carbohydrate (161·52 kcal), 9·84g of fat (88·5 kcal)). Six of these participants returned for a second visit and followed the same protocol to allow the repeatability of the measurements to be determined. The other six participants returned for a second visit during which they followed the same protocol but consumed water instead of milkshake, to allow the effects of gut distension to be distinguished from the response to macronutrients.

**Using ultrasound to investigate changes in blood supply to the small intestine following stress.** Stress is known to reduce blood flow to the intestine[16,17]. To determine the utility of ultrasound in measuring the response of the small intestinal microvasculature to physical stress, 4 participants attended the clinical research unit for a single visit where they were imaged while being fasted and 20 min after the ensure intake. Then, they placed their hands in cold water (4·5° - 6·5°) for 1 to 2 minutes, to cause mild physical stress[18], and acquisitions were taken.

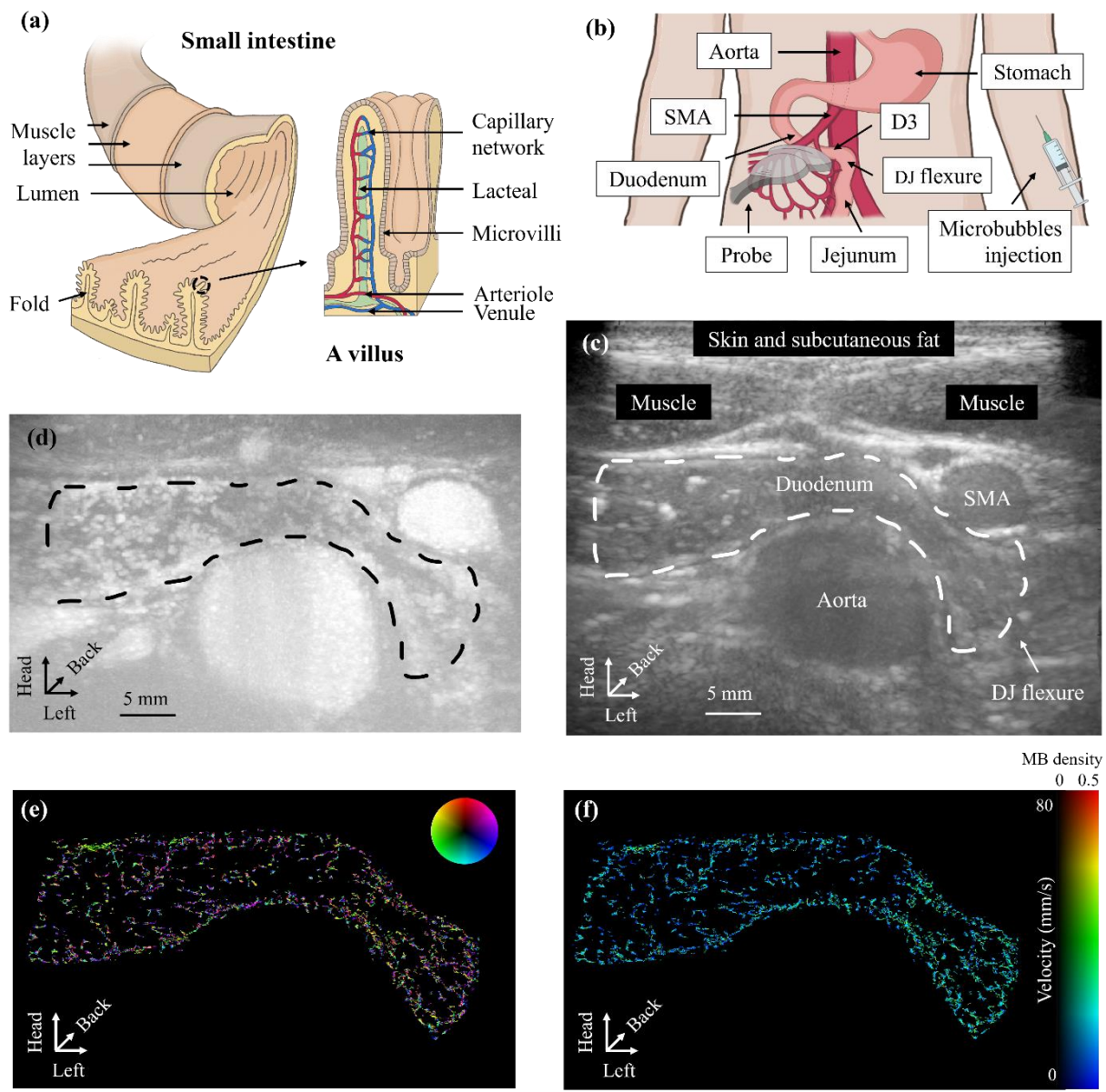

**Figure 1: Experimental set-up and representative CEUS and SRUS/ULM acquisitions.**
(a) Schematic diagram illustrating the structure of the small intestine, focusing on villus microvascular structure. (b) Schematic diagram showing the positioning of the ultrasound probe on the human volunteer to image section D3 of the duodenum, with surrounding organs used as anchor points. SMA: superior mesenteric artery, DJ: duodenojejunal. (c) Maximum intensity projection (MIP) of a B-mode ultrasound view of section D3 of the human duodenum and surrounding organs, when the subject was supine. Dynamic rage: [-60 0]. (d) Maximum intensity projection (MIP) of contrast enhanced data corresponding to the above image of the duodenum. Dynamic rage: [-60 0]. (e) Example of super-resolution (SR) angle map of this image of the duodenum after processing. (f) Example of super-resolution (SR) speed map of this image of the duodenum after processing.

**2D Ultrasound acquisition**

CEUS data were acquired using a GEL3-12D linear probe connected to a Vantage 256 research system (Verasonics, Kirkland, USA) at a central frequency of 5 MHz. A bolus of 1·6 ml Sonovue microbubbles was injected over a period of 20 seconds followed by a 1 ml saline flush. A 3 pulse amplitude modulation (AM) sequence (half-full-half), with 4 angles, was used for the acquisitions and the compounded frame rate was 300 Hz. 5 seconds data were acquired for 5 minutes after the bolus injection. A mechanical index of 0·05 was used to avoid the destruction of the contrast agents.

**3D Ultrasound acquisition**

Data were acquired using a matrix array probe with 1024 elements at a central frequency of 7·8 MHz (Vermon, France) connected to another Verasonics Vantage 256 system. Both 2D and 3D data were acquired alternatively by switching two probes and their corresponding systems between each acquisition. A bolus of 2·5 ml SonoVue microbubbles was injected over a period of 30 seconds followed by a 1 ml saline flush, to allow for sufficient bubbles for the matrix array.

**Data Processing**

All processing was conducted using MATLAB (R2023a, MathWorks, Natick, MA, USA). All the data were first beamformed using cross-angular delay multiplying and sum beamforming (CADMAS)[19]. For each participant, only data with no strong gut movement and no visible gas within the gut lumen were included for post-processing, avoiding large and complex motion and artefacts from the strong reflection of gas in the gut. As the participants were fasted, only a few had notable gas bubbles, and there were no imaging sessions when gas bubbles were present for the whole of the acquisition. To allow comparison of the data between visits and subjects, the data selected for analysis was taken from approximately the same time period following the bolus. Frame selection was used to discard frames with too high gut motion, such as in-and-out of plane motion, strong peristalsis, or with too much probe motion. This selection was based on the cross-correlation between two successive frames and on visual evaluation; each selected period accounts for about 1 second.

For each selected period, data with slow- and fast-moving bubbles were separated based on their speed using a low pass and high pass Butterworth filter. The cut-off frequency was set up to 32·5 Hz, corresponding to a speed of 10 mm/s[20–23]. The tissue signal was extracted from the B-Mode acquisition using a clutter filter (singular value decomposition)[24] to remove residual MBs signal to carry out a two-stage motion estimation[25]. The estimation was then applied on each of the period of contrast data to correct the motion. A second two-stage motion estimation on the tissue image was performed to estimate the motion between each period and reconstruct the slow, fast and full datasets.

The CEUS data generated at this stage were used for contrast quantification. Selected CEUS data with good enough image quality were further processed to generate SR maps. The SR processing pipeline was applied separately to the slow- and fast- moving bubbles to produce more accurate tracking. For these data, noise reduction was performed using a 3D median filter in spatial and temporal dimensions followed by an automatic adaptive local noise thresholding.

A region of interest (ROI) was drawn around the gut to perform SR only in the masked area. Then, localisation and tracking were done[6], generating slow- and fast-moving bubbles SR maps. The two maps were finally combined to generate complete SR maps (Figure 2).

**Quantification of intestinal perfusion**

**CEUS Data Quantification:** After the motion correction, CEUS data were used for quantification. For each dataset, a region of interest (ROI) was drawn around the gut, by an individual trained by a clinician. The average intensity over time was calculated, followed by an Otsu threshold to remove background and retain only the signal from the vasculature within the ROI. Subsequently, the intensity of the pixels was summed and divided by the length of the ROI, to allow fairer comparison within a participant and across participants. This provides a number that serves as a proxy for the amount of blood flowing in this section of the intestine during the time of the acquisition.

**SR Data Quantification:** To visualise and assess the microcirculation, slow SR maps were used for the SR data quantification. Only bubbles tracks over more than 10 successive frames and with an average speed lower than 10 mm/s were kept, in order to capture information from the microvasculature but not from the larger vessels in the gut wall[20–23].

**Statistical analysis**

Data is shown as mean ± standard error of the mean. Student's paired t-tests were used to compare the different states (fasted, after nutrient or water ingestion, after stress). No power calculation was performed as there was no prior data to base this on.

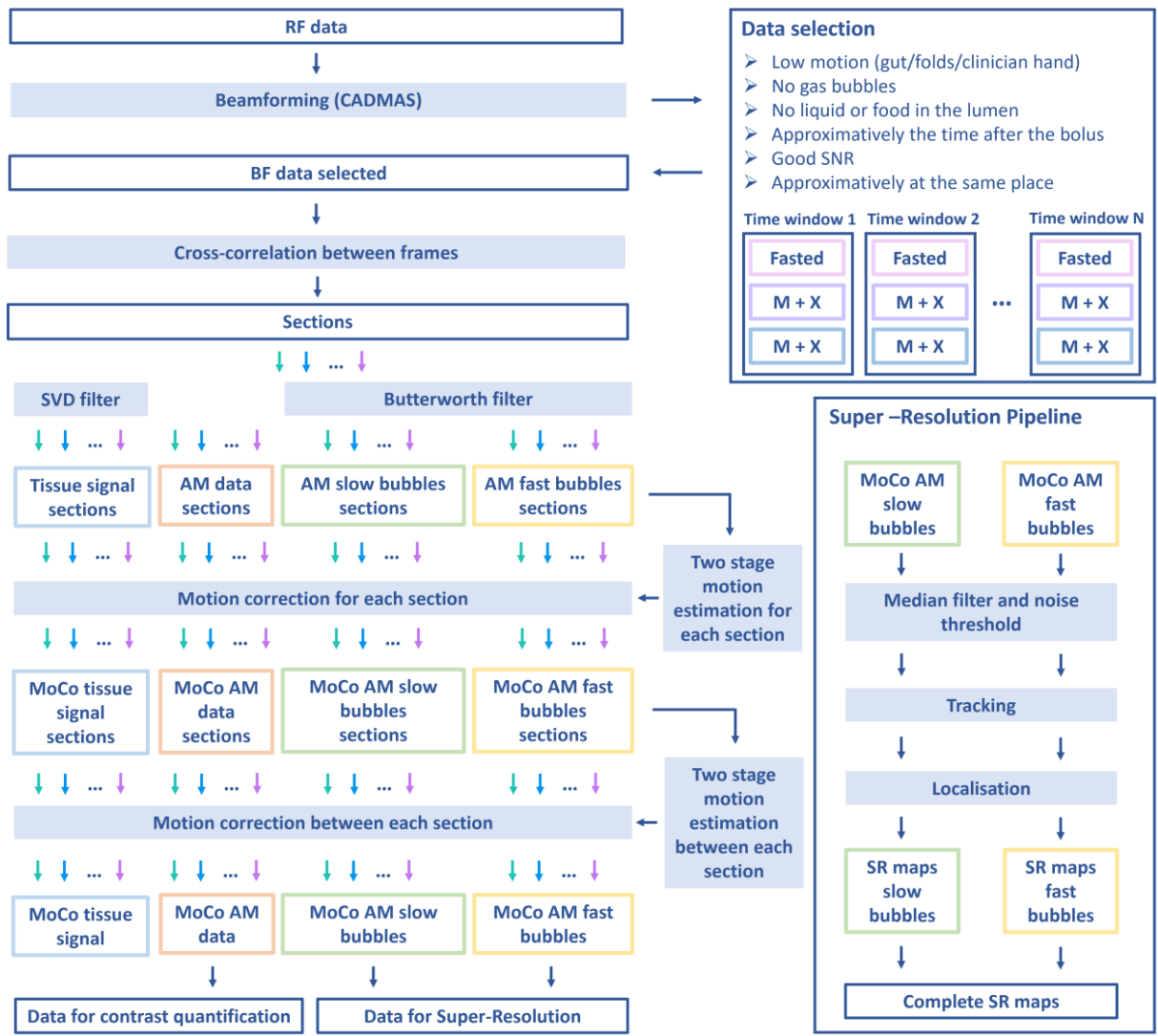

**Figure 2: Schematic diagram outlining the data processing pipeline.** RF: radiofrequency, BF: beamformed, CADMAS: cross-angular delay multiplying and sum, SNR: signal-to-noise ratio, M: milkshake, M+X where X represent a time in seconds, after the injection of the bolus of microbubbles, SVD: singular value decomposition, AM: amplitude modulation, MoCo: motion corrected, SR: super-resolution.

## RESULTS

**Participant demographics**

There were an equal number of male (10) and female (10) volunteers, aged between 19 and 31 years old (25·7 +/- 0·8), with an average BMI of 21·3 +/- 0·3. Ethnicity: East Asian (10), White (7), Latino (2), South Asian (1).

**Microcirculation inside the human small intestine.**

SRUS allowed us to visualize the microcirculation within the human small intestine. Figure 3 (a) shows a SR map of the fast bubble flow and slow bubble flow in a section of jejunum in a single participant generated from an acquisition containing a relatively high concentration of bubbles. Some parts of the gut which are more vascularized are highlighted with white arrows and appear to correspond to the plicae circulares, folds expected on the luminal surface of the jejunum. Higher magnification reveals that these areas of faster flow are connected to vessels with a slower flow which spread from the edges of the denser vascularized areas. These are highlighted in the insets by the yellow arrows. Structurally, this seems to correspond to the capillary networks of the villi. At lower bubble concentrations it is more difficult to discern the macrostructure of the gut, but these acquisitions lend themselves well to extracting information on individual bubbles to plot their individual paths (Figure 3 (b)).

**Estimating the blood flow speed of the small intestine capillaries.**

The single slow-moving bubbles extracted from the data were assumed to be from within the capillary network of the villi. From the slow-moving SR map, a first selection was made to keep only the bubbles tracked for over 75 frames, representing 0·25 seconds, and with an average speed lower than 10 mm/s, allowing for the extraction of arterioles and pre-capillaries as well as capillaries (speed distribution in Figure 3 (c)). The tracks are localised sparsely inside the gut, and not aggregated around a specific place, in accord with the structure of the villi (Supplementary video 1). To evaluate capillary speed more accurately, a second selection was made to remove tracks thought to represent the movement of bubbles specifically within arterioles or venioles. Only tracks with a speed lower than 5 mm/s and tracked for more than 10 consecutives frames were kept (speed distribution in Figure 3 (d))[20–23]. In total, 631 MBs were extracted. The average distance between the beginning and the end of the tracks was 0·46 ± 0·01 mm (mean ± SEM) in accord with a villus height of approximately 1 mm[26]. The mean speed was 2·78 ± 0.05 mm/s (mean ± SEM).

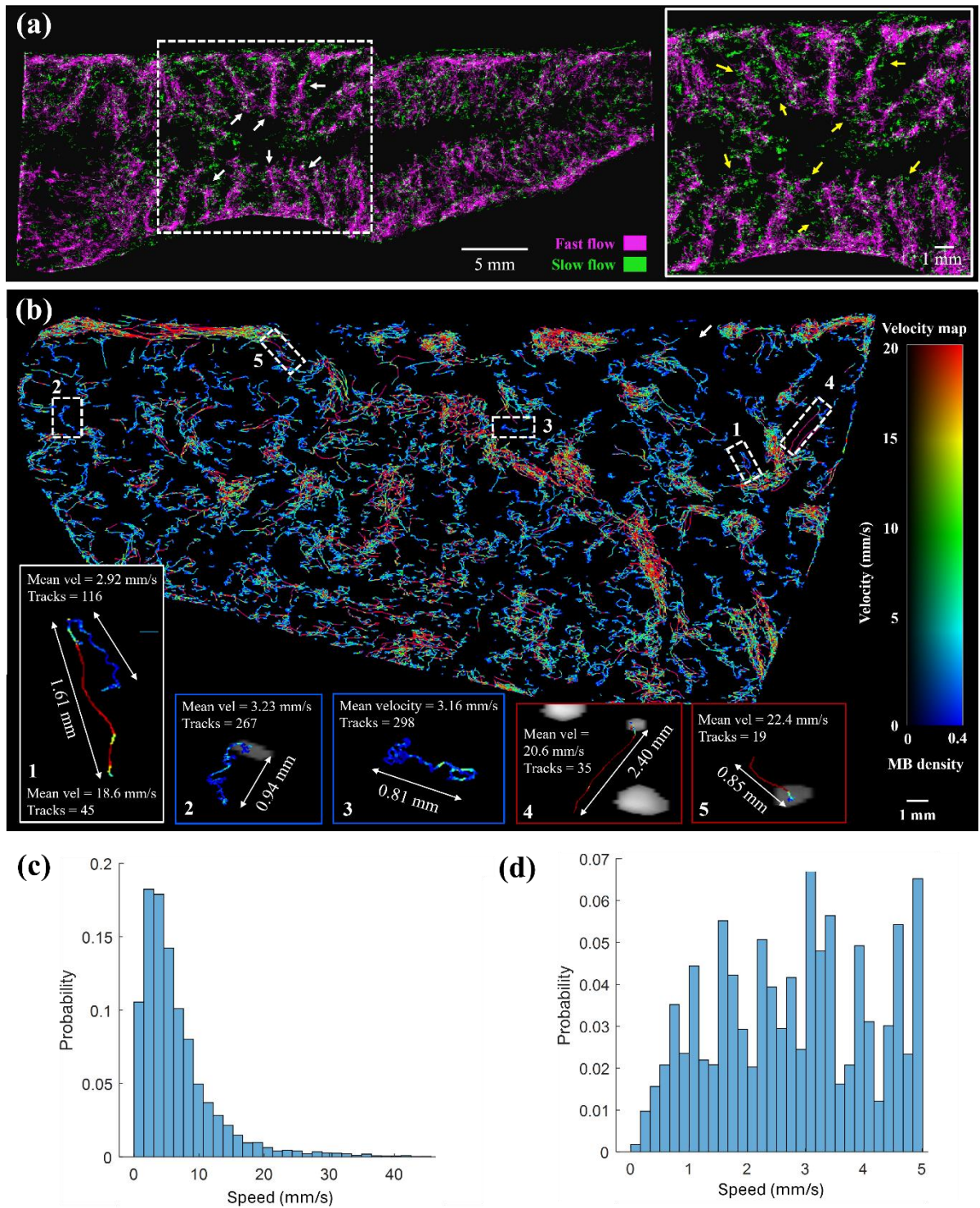

**Figure 3: Determining the microvasculature of the human small intestine.** (a) SR density map of a jejunum section with separated fast and slow flow, highlighting the plicae circulares and the villi location. (b) SR velocity map of another jejunum section with highlighted individual bubbles tracks, showing slow (insets 2 and 3) and fast (insets 4 and 5) moving bubbles, and a single fast bubble slowing down, likely entering a smaller vessel, and then moving slowly through it (inset 1). (c) Speed distribution of the slow-moving bubbles; the average speed was 6.40 ± 0.24 mm/s (mean ± SEM). (d) Speed distribution of the slow-moving

bubbles after selecting tracks with a speed lower than 5 mm/s; the average speed was 2.78 ± 0.05 mm/s (mean ± SEM).

## 3D imaging

3D acquisitions of the jejunum were taken in two participants. Figure 4 shows a 3D acquisition of the jejunum. The dotted boxes pink and blue show the same region on different slices that appears to reflect the structure of an intestinal fold.

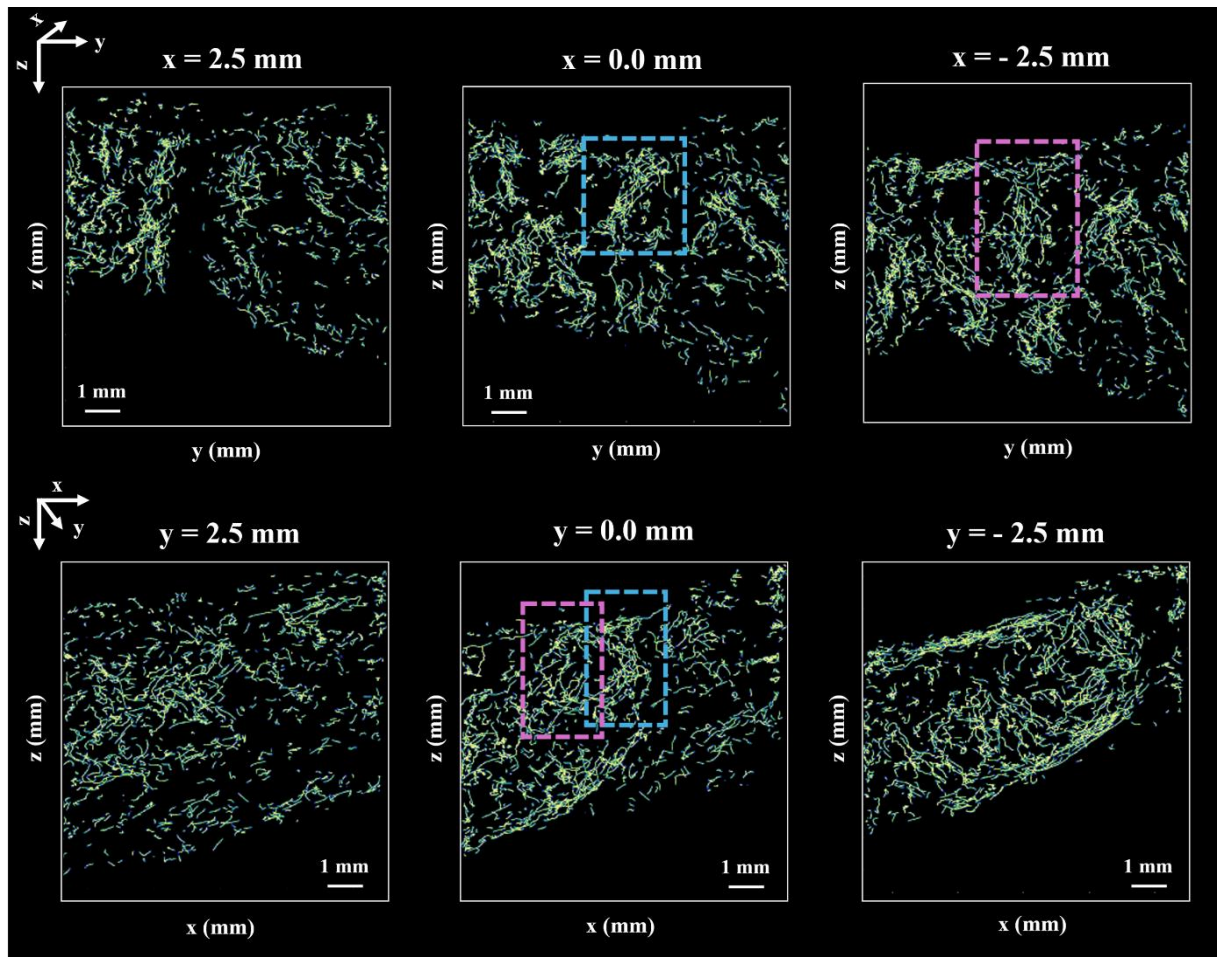

**Figure 4: 3D SR density map of the microvasculature of the human jejunum.** The dotted boxes pink and blue show the same region on different slices that appears to reflect the structure of an intestinal fold. Slab thickness: 3 mm.

## Contrast data quantification

Figure 5 (a) shows representative CEUS acquisitions from a single participant taken in the fasted state, 5 min and 20 min after the milkshake ingestion. The imaging plane was relatively similar in the three acquisitions, the shifts being necessary to follow the displacement of the duodenum. The ingestion of nutrients significantly (p = 0·04) increased the blood flow to the duodenum by an average of 36·2 ± 12·2 % (Figure 5 (b)). Further analysis of the timeline of perfusion following milkshake or water ingestion suggested that the milkshake resulted in higher perfusion from 5 to 30 minutes, though these changes did not achieve statistical significance at individual time points (Figure 5 (c)). To investigate if the increase of perfusion following food consumption was mechanically driven by distension of the gut, 6 participants consumed the same volume of water instead of the milkshake on their second visit. Water intake resulted in a small, non-statistically significant increase in perfusion of the duodenum (Figure 5 (d)). Perfusion after Milkshake (visit 1) or water (visit 2) intake were then compared for the same 6 participants, following normalization to the corresponding fasted state, demonstrating that milkshake ingestion significantly increased perfusion compared to water ingestion (Figure 5 (e)). Perfusion of section D3 of the duodenum in response to acute stress was also evaluated in two participants, both showing a decreased perfusion (Figure 5 (g)).

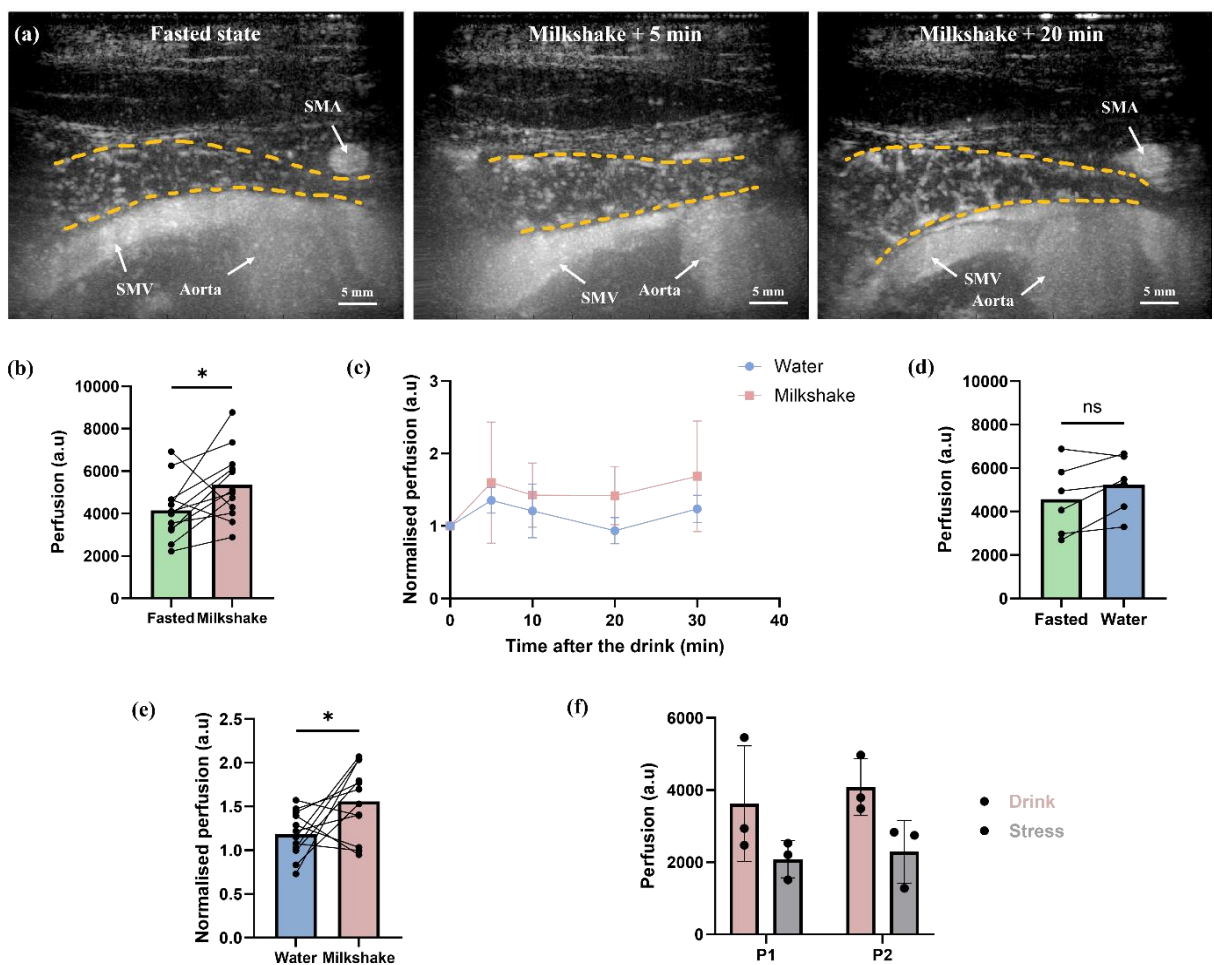

**Figure 5: Determining the effect of food ingestion on vascular perfusion of the human duodenum.** (a) MIP of 1 second ultrasound contrast data of the duodenum for the same participant in three different states: fasted, 10 min after the milkshake, and 20 min after the milkshake. Dynamic range: [-35 0] dB. Yellow dotted lines show the duodenum delimitation. SMA: superior mesenteric artery, SMV: superior mesenteric vein. (b) Fasted vs milkshake for the visit 1 of participants 1-12, p-value = 0.0421. (c) Evolution of the perfusion inside the duodenum through time for participant 1-12. Error bars show the standard deviation. (d) Fasted

vs water for the visit 2 of participants 7-12, p-value = 0.0698. (e) Water (visit 2) vs milkshake (visit 1) for participants 7-12, p-value = 0.0408. (f) Perfusion of section D3 of the duodenum in response to acute stress for two participants (17 and 19). Each point represents a measure of perfusion at different times over 2 minutes.

**DISCUSSION**

This work has demonstrated the feasibility of non-invasive *in vivo* ultrafast CEUS and SRUS imaging of the microvasculature of the human small intestine. For the first-time, high-resolution images of the small intestine microvasculature were generated, and CEUS was able to detect the changes in perfusion that occurred following food ingestion.

This study demonstrates that human duodenum and jejunum micro-vessels with flow as slow as below 1 mm/s, which is consistent with that of capillaries, can be detected and directly visualised, using a non-invasive transabdominal approach. This has never been achieved as far as we are aware. Though a number of studies have used different approaches to assess capillary blood flow speed in different organs and species[20–23,27], to our knowledge, only one published study has achieved this in the capillaries of human villi using an endoscopic approach, estimating capillary blood flow speed at 0·57 mm/s in duodenal capillaries in healthy paediatric subjects, using probe-based confocal laser endomicroscopy[20]. Overall, speed flow in capillaries is generally estimated to be lower than 1 mm/s[23] but can sometimes be higher, for example in muscles (1·14 mm/s)[27]. It is important to note that blood flow speed inside capillaries varies continuously depending on the state of the organ and that blood may move episodically due to vasomotion. Differentiating capillaries from arterioles or venules is challenging and, in our case, can only be based on the vessels speed for which there are no definitive values. To select capillaries, the choice was made to set a threshold on the speed of the tracks (5 mm/s). This threshold ensures that all capillaries were selected, but some pre-capillaries and arterioles were likely also included in the calculation of the average speed, which may explain the relatively high values obtained.

The results obtained on CEUS quantification were in accord with expectations regarding blood flow following food ingestion or acute stress[2,15,28]. It has previously been suggested that individual villi are not all perfused in the fasted state, prioritising blood flow to other organs (2,3). Our observations support this and suggest that blood flow is limited in many of the villi before food consumption and that perfusion increases following the food consumption. Our observations also suggest that mechanical stimulation, in this case by ingestion of water, may cause a small increase in duodenal blood flow, but this increase is lower than that invoked by the same volume of nutrient-containing fluid. Preliminary work also seems to show a trend of reduced blood flow under acute stress. The variability in the data was likely driven by several factors. These include the difficulty in imaging repeatedly the same region of the intestine in the exact same plane, the fact that the pressure exerted by the probe may affect blood flow, as well as the natural variation that would be expected in any physiological process. However, this pilot study demonstrate that CEUS can detect physiological effects on the small intestine perfusion.

**Limitation of the study**

Gut motion includes unpredictable local movements of the folds and villi and longer and more coordinated peristaltic movements, both inducing strong out of plane motion (Supplementary Video 2). High-resolution images could only be captured when the gut showed little or no motion, and for short periods of approximately 1 second, as it was unlikely that the gut would remain motionless for longer periods. These short acquisitions and the gut motion made it challenging to produce a vascular map of the gut. The development of better adapted motion correction algorithms for these specific cases may overcome the limitations of movement. It is also possible that drugs to reduce gut motion could be taken before imaging to identify structural changes, though this could also confound the results obtained. 3D imaging could also address the out of plane motion issue. We carried out 3D imaging on two participants during the study and found it to be feasible in both. There is evidence showing that gut motility tends to be lower in IBD patients[29,30], which might make it easier to apply it to these patients for diagnostic and prognostic purposes.

Distortion of the ultrasound wave due to the presence of abdominal fat above the targeted section of the small intestine, and especially the duodenum, also presented a challenge. For some images the loss of resolution of the contrast agents inside the gut complicates the isolation and tracking of microbubbles. However, superficial sections of the jejunum allowed high quality SR images of the gut and the visualisation of slow-moving bubbles flowing in what are believed to be the capillaries of the villi.

In summary, this work has demonstrated the feasibility of 2D SRUS/ULM for imaging the microcirculation and evaluating the blood flow speed of the capillaries in the human small intestine. 3D SRUS/ULM has also been proven possible. Ultrafast CEUS appears to be a potentially useful tool for assessing perfusion of the duodenum and detecting changes in blood flow. In the future, these techniques may become useful in diagnosis and monitoring of gut disease and the study of gut physiology.

**ACKNOWLEDGMENTS**

The authors would like to thank the Imperial Clinical Research Facility (ICRF) for hosting the clinical acquisitions.

**COMPETING INTERESTS**

The authors declare no competing interests.

**FUNDING**

CV is supported by the Engineering and Physical Sciences Research Council (EPSRC) Centre of Doctoral Training (CDT). MT and KGM are both supported by Diabetes UK (DUK) (20/0006295), the BBSRC (BB/W001497/10) and the Wellcome Trust (310835/Z/24/Z). KGM is additionally supported by the DUK (18/0005886), MRC (MR/Y013980/1) and BBSRC (BB/X017273/1). GF is supported by the BBSRC (BB/X011054/1, BB/X017273/1) and Nxera

Pharma UK. CS is supported by the International Human Frontier Science Program Organization (LT0036/2022-L). MaT and JY are supported by the National Institute for Health Research i4i (NIHR200972). SY is supported by the BBSRC (BB/W001497/10).

**REFERENCES**

1. Matheson, P. J., Wilson, M. A. & Garrison, R. N. Regulation of intestinal blood flow. *Journal of Surgical Research* **93**, 182–196 (2000).

2. Granger, D. N., Holm, L. & Kvietys, P. The gastrointestinal circulation: Physiology and pathophysiology. *Compr Physiol* **5**, 1541–1583 (2015).

3. Christensen-Jeffries, K., Browning, R. J., Tang, M. X., Dunsby, C. & Eckersley, R. J. In vivo acoustic super-resolution and super-resolved velocity mapping using microbubbles. *IEEE Trans Med Imaging* **34**, 433–440 (2015).

4. Errico, C. *et al.* Ultrafast ultrasound localization microscopy for deep super-resolution vascular imaging. *Nature* **527**, 499–502 (2015).

5. Christensen-Jeffries, K. *et al.* Super-resolution Ultrasound Imaging. *Ultrasound in Medicine and Biology* vol. 46 865–891 Preprint at https://doi.org/10.1016/j.ultrasmedbio.2019.11.013 (2020).

6. Yan, J., Zhang, T., Broughton-Venner, J., Huang, P. & Tang, M. X. Super-Resolution Ultrasound Through Sparsity-Based Deconvolution and Multi-Feature Tracking. *IEEE Trans Med Imaging* **41**, 1938–1947 (2022).

7. Harput, S. *et al.* Two-Stage Motion Correction for Super-Resolution Ultrasound Imaging in Human Lower Limb. *IEEE Trans Ultrason Ferroelectr Freq Control* **65**, 803–814 (2018).

8. Demené, C. *et al.* Transcranial ultrafast ultrasound localization microscopy of brain vasculature in patients. *Nat Biomed Eng* **5**, 219–228 (2021).

9. Huang, C. *et al.* Super-resolution ultrasound localization microscopy based on a high frame-rate clinical ultrasound scanner: An in-human feasibility study. *Phys Med Biol* **66**, (2021).

10. Zhu, J. *et al.* Super-Resolution Ultrasound Localization Microscopy of Microvascular Structure and Flow for Distinguishing Metastatic Lymph Nodes - An Initial Human Study. *Ultraschall in der Medizin* **43**, 592–598 (2022).

11. Denis, L. *et al.* Sensing ultrasound localization microscopy for the visualization of glomeruli in living rats and humans. *EBioMedicine* **91**, (2023).

12. Yan, J. *et al. Transthoracic Super-Resolution Ultrasound Localisation Microscopy of Myocardial Vasculature in Patients*.

13. Porte, C. *et al.* Ultrasound Localization Microscopy for Breast Cancer Imaging in Patients: Protocol Optimization and Comparison with Shear Wave Elastography. *Ultrasound Med Biol* **50**, 57–66 (2024).

14. Morris, M. *et al.* In-Patient Repeatability and Sensitivity Study of Multi-Plane Super-Resolution Ultrasound in Breast Cancer. Preprint at https://doi.org/10.1101/2024.10.15.24315514 (2024).

15. Price, H. L., Cooperman, L. H. & Warden, J. C. *Control of the Splanchnic Circulation in Man ROLE OF BETA-ADRENERGIC RECEPTORS*. http://ahajournals.org.

16. Perko, M. J. *et al. Mesenteric, Coeliac and Splanchnic Blood Flow in Humans during Exercise*. *Journal of Physiology* (1998).

17. Otte, J. A., Oostveen, E., Geelkerken, R. H., Groeneveld, A. B. J. & Kolkman, J. J. Exercise induces gastric ischemia in healthy volunteers: a tonometry study. *J Appl Physiol* **91**, 866–871 (2001).

18. Sramek, P., Simeckova, M., Jansky, L., Savlikova, J. & Vybiral, S. Human physiological responses to immersion into water of different temperatures. *Eur J Appl Physiol* **81**, 436–442 (2000).

19. Smith, C. A. B. *et al.* Enhanced Ultrasound Image Formation with Computationally Efficient Cross-Angular Delay Multiply and Sum Beamforming. Preprint at https://doi.org/10.1101/2024.09.03.611015 (2024).

20. Zaidi, D. *et al.* Capillary Flow Rates in the Duodenum of Pediatric Ulcerative Colitis Patients Are Increased and Unrelated to Inflammation. *J Pediatr Gastroenterol Nutr* **65**, 306–310 (2017).

21. Nakajima, Y., Baudry, N., Duranteau, J. & Vicaut, E. Microcirculation in Intestinal Villi. *Am J Respir Crit Care Med* **164**, 1526–1530 (2001).

22. Ruh, J. *et al. Measurement of Blood Flow in the Main Arteriole of the Villi in Rat Small Intestine with FITC-Labeled Erythrocytes 1*. *Microvascular Research* vol. 56 (1998).

23. Chan, C. W., Leung, Y. K. & Chan, K. W. Microscopic anatomy of the vasculature of the human intestinal villus-a study with review. *European Journal of Anatomy* **18**, 291–301 (2014).

24. Demené, C. *et al.* Spatiotemporal Clutter Filtering of Ultrafast Ultrasound Data Highly Increases Doppler and fUltrasound Sensitivity. *IEEE Trans Med Imaging* **34**, 2271–2285 (2015).

25. Yan, J. *et al.* Transthoracic ultrasound localization microscopy of myocardial vasculature in patients. *Nat Biomed Eng* **8**, 689–700 (2024).

26. Gannon, B. J. & Perry, M. A. Histoanatomy and ultrastructure of vasculature of alimentary tract. in *Comprehensive Physiology* 1301–1334 (Wiley, 1989). doi:10.1002/cphy.cp060136.

27. Ivanov, K. P., Kalinina, M. K. & Levkovich, Y. I. *Blood Flow Velocity in Capillaries of Brain and Muscles and Its Physiological Significance*.

28. Leigh, S.-J. *et al.* The impact of acute and chronic stress on gastrointestinal physiology and function: a microbiota-gut-brain axis perspective. *J Physiol* **601**, 4491–4538 (2023).

29. Bassotti, G. *et al.* Abnormal gut motility in inflammatory bowel disease: an update. *Techniques in Coloproctology* vol. 24 275–282 Preprint at https://doi.org/10.1007/s10151-020-02168-y (2020).

30. Yung, D. *et al.* Morpho-functional evaluation of small bowel using wireless motility capsule and video capsule endoscopy in patients with known or suspected Crohn's disease: pilot study. *Endosc Int Open* **04**, E480–E486 (2016).

# SUPPLEMENTARY VIDEO

**Supplementary Video 1:** Video highlighting specific steps for the extraction of slow-moving bubbles and tracking of an individual slow-moving bubble. The bubble generally moves slowly, staying approximately at the same place, likely reflecting them entering the capillary network of a villus.

**Supplementary Video 2:** Video showing examples of gut motion. On the left side, coordinated contraction of the muscles of the gut wall (peristalsis) and wave motion of the folds and villi. On the right side, an acquisition taken on the same participant, later in the bolus (lower bubble concentration), with little gut motion.